\documentclass[a4paper,11pt,twosided,fleqn,epsfig]{article}

\usepackage{latexsym}

\usepackage{graphicx}

\usepackage{amssymb}
\usepackage{amsmath}
\usepackage{amsfonts}
\usepackage[usenames,dvipsnames]{color}

\begin{document}

\title{\bf A finite-time exponent for random Ehrenfest gas }
\author{Sanjay Moudgalya$^1$, Sarthak Chandra$^1$, Sudhir R. Jain$^2$\footnote{Email: srjain@barc.gov.in; Tel.: +912225593589}\\
$^1$\textit{Indian Institute of Technology, Kanpur 208016, India}\\
$^2$\textit{Nuclear Physics Division, Bhabha Atomic Research Centre,}\\ {\it Mumbai 400085, India} }
\date{}
\maketitle
\normalsize
\begin{abstract}
We consider the motion of a system of free particles moving on a plane with regular hard polygonal scatterers arranged in a random manner. 
Calling this the Ehrenfest gas, which is known to have a zero Lyapunov exponent, we propose a finite-time  exponent to characterize its dynamics. 
As the number of sides of the polygon goes to infinity, when polygon tends to a circle, we recover the usual Lyapunov exponent for the Lorentz gas from the exponent proposed here. 
To obtain this result, we generalize the reflection law of a beam of rays incident on a polygonal scatterer in a way that the formula for the circular scatterer is recovered in the limit of infinite number of vertices. Thus, chaos emerges from pseudochaos in an appropriate limit. 
\end{abstract}

\newpage
\section{Introduction}

Is macroscopic stochasticity originating from microscopic chaos ? This question underlies a fundamental connection between deterministic chaos and statistical mechanics. Krylov \cite{krylov} argued the case for mixing dynamics as a requirement for statistical laws to hold. A connection between transport property and chaotic microscopic motion was proposed on the base of an experiment more than fifteen years ago \cite{gaspard1998}. Considering the Brownian movement of a colloidal particle in water, the authors concluded that the sum of positive Lyapunov exponents is positive. Since Lyapunov exponents measure whether the dynamics sensitively depends on the initial conditions, this quantity seemed to present for the first time a direct connection of microscopic dynamical instability to relaxation properties of a many-body system, propounded in the classic work of Krylov \cite{krylov}. Krylov had laid the dynamical foundations of Lorentz gas and estimated the Lyapunov exponent for this system. An enormous amount of work has been carried out since then, summarized in reviews and books \cite{ehh,HvB-RMP,gaspardbook,cpd2000,cpd2014}.
   
Subsequent to the work by Gaspard et al. \cite{gaspard1998}, there has been a vibrant debate where, mainly, at the centre of the debate is the gas with polygonal scatterers. We call this the Ehrenfest gas to distinguish it from the Lorentz gas where the scatterers are circular in shape. Due to the absence of curved segments of the scatterers in the Ehrenfest gas, there is no exponential separation of infinitesimally separated trajectories and thus the Lyapunov exponent is zero. Dettmann et al. \cite{henk} argued that while an estimate of the Kolmogorov-Sinai entropy was given in \cite{gaspard1998},  based on time series recorded from positions of a colloidal particle moving in a fluid; very similar results were obtained for certain models with strictly zero Lyapunov exponent. Numerical experiments on the wind-tree model confirmed these arguments \cite{henk}.  A mixture of normal and anomalous diffusive behaviour of point particles moving along a polygonal channel was observed in careful numerical experiments by Alonso et al. \cite{ar}. Thus,  at the microscopic level, it is not clear what degree of non-integrability \cite{footnote} is necessary to ensure typical transport properties \cite{spohn,grassberger,gaspard1999}. 

For calculating the Lyapunov exponent of the Lorentz gas, Sinai and Dorfman \cite{krylov-sinai,bob} have considered a narrow beam of two rays separated by an  infinitesimal angle $\alpha$ and calculated the typical time evolution of the separation. Suppose the ends of the rays are a part of a circular wavefront of  radius $\rho_-$ before a collision, one may use the well-known mirror formula for reflection off a concave surface to determine the radius $\rho_+$ of the wavefront after collision. This result is crucial for their calculation of the Lyapunov exponent. In calculating the Lyapunov exponent, as mentioned above, the beam is considered infinitesimally narrow so that its bounding rays suffer collision with the same set of scatterers arranged in a periodic manner. For the Ehrenfest gas, this can also be achieved. However, there appears a time during the evolution when a part of the beam falls on one side of an $N$-sided polygonal scatterer, whereas the other part strikes an adjacent side. This happens for a beam that is not infinitesimal, and, for scatterers with $N$ large. Here we provide a definition of a finite-time exponent for pseudochaotic systems like the Ehrenfest gas, by taking into account the relation between curvatures before and after the beam splits. To summarize, there are two main results we present - a collision rule to quantitatively describe beam splitting, and, an exponent incorporating this rule in the formulation due to Sinai \cite{krylov-sinai}, with an estimate of the time over which this exponent may be used. 

We restrict ourselves to a regime of low density of randomly placed scatterers as analytic estimates become too hard for higher densities. As the number of sides of the polygonal scatterers increases, the expressions found by us recover the well-known results for the circular scatterers. Thus, in a sense, chaos emerges from pseudochaos.  

We recall that finite-time Lyapunov exponents are defined for chaotic systems also but the system we are dealing with here is a non-chaotic, non-integrable one (we have called this pseudo-chaotic) \cite{footnote}. 

\section{Collision rule}
\begin{figure}[!]
\begin{center}
\includegraphics[scale=1.0,height=66mm]{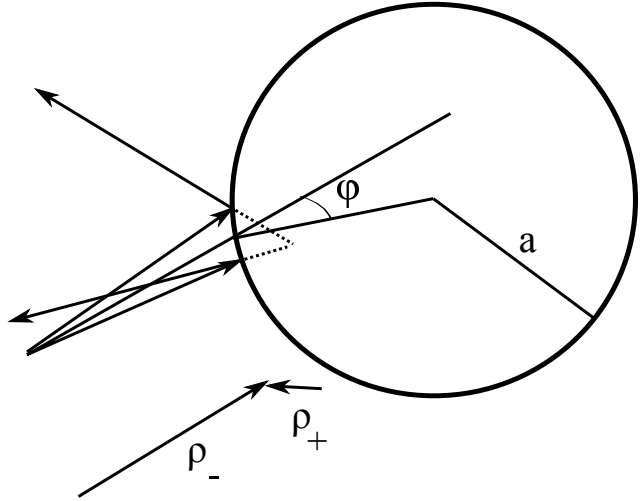}
\end{center}
\caption{A beam of rays is incident on a disc of radius `a' scatters such that the  radius of curvature of the wavefront corresponding to them changes from $\rho _-$ to $\rho _+$.}
\label{fig:circref}
\end{figure}

In the case of circular scatterers (Lorentz gas), the relation between the radius of curvature before and after the collision is given by the well-known reflection formula for spherical mirrors (Fig.~\ref{fig:circref}):  
\begin{equation}\label{eq:circref}
\frac{1}{\rho _+} = \frac{1}{\rho _-} + \frac{2}{a\cos \varphi }.
\end{equation}
Note that $\rho_-$ is usually very large in a Lorentz gas of low density, and hence $\rho_+$ is approximately $a \cos{\varphi}/2$. 
\begin{figure}[!]
\begin{center}
\includegraphics[scale=1.0,height=66mm]{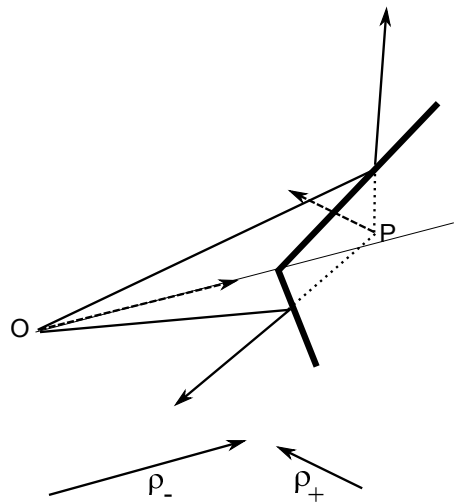}
\end{center}
\caption{A beam of rays incident about a vertex of a polygonal scatterer, where the radius of curvature of the wavefront changes from $\rho _-$ to $\rho _+$. The dashed arrows from O and P respectively denote the radii of curvatures of the beam before and after the reflection. }
\label{fig:polyref}
\end{figure}

\begin{figure}[!]
\begin{center}
\includegraphics[scale=1.0,height=66mm]{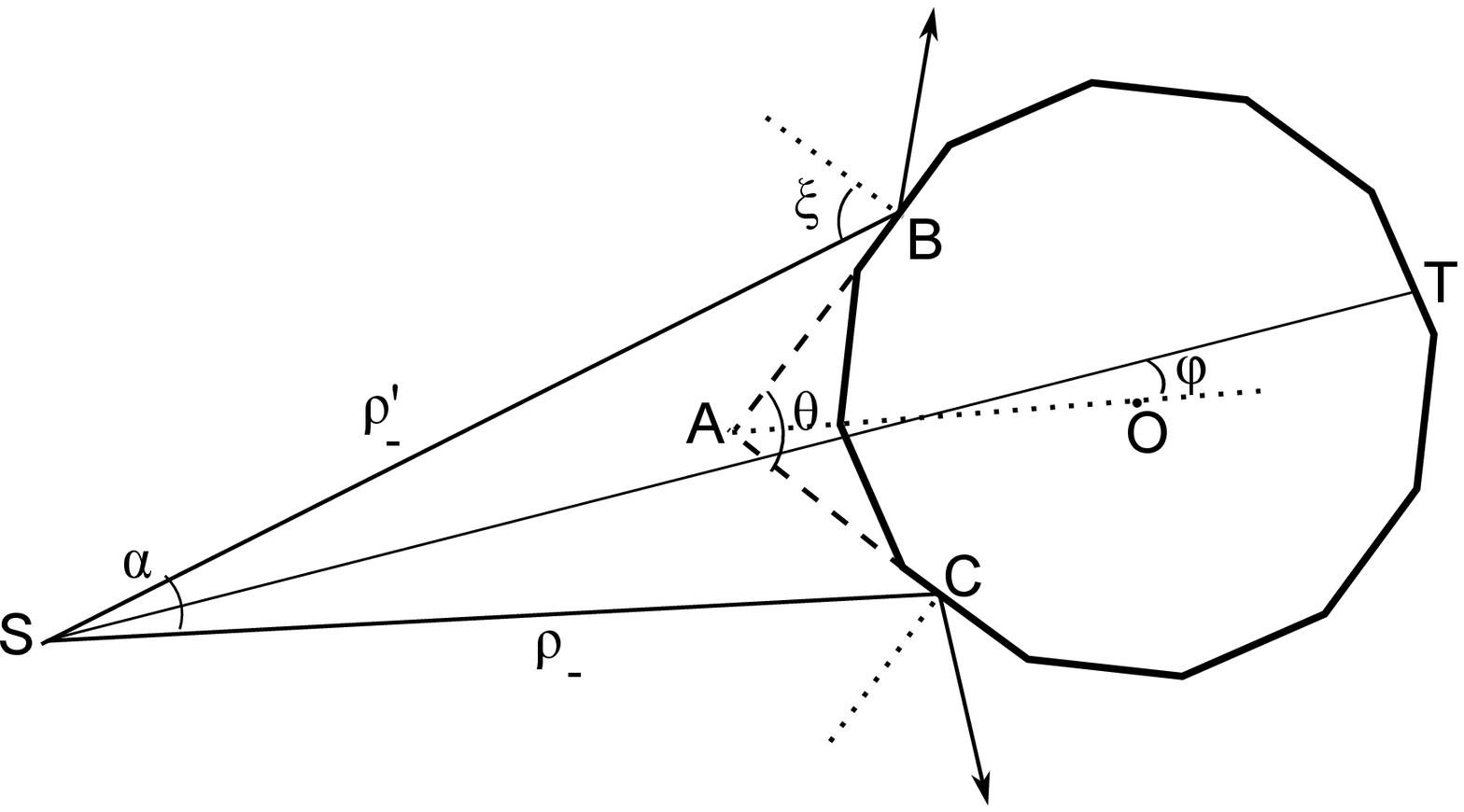}
\end{center}
\caption{Geometry of reflection off a 12-sided polygonal scatterer. A beam of rays with angular width $\alpha $ is incident such that its two ends strike two distinct sides of the scatterer. In the case shown, there are three vertices enclosed by the beam, with point A as the point where the two sides (when extended) meet. The length from A to where the rays hit at B and C are $s_1$ and $s_2$ respectively. The perpendicular distance between the bisectrix of the ray and the centre of the scatterer $O$, is $y$. The change in radius of curvature upon such a reflection is given by Eq.~(\ref{eq:polyref}). The orientation of the beam with respect to the scatterer is given by the angle $\varphi$, an angle between the incident beam and OA. } 
\label{fig:fig3}
\end{figure}

For the reflection off a polygonal scatterer as shown in Fig.~\ref{fig:polyref}, the formula requires a generalization of Eq.~(\ref{eq:circref}).
We first consider the case in which the rays hit a polygonal scatterer with a single vertex between them, as shown in Fig.~\ref{fig:polyref}.
The geometry of the reflection off a polygon is shown in Fig.~\ref{fig:fig3}.
The beam that hits the scatterer is characterized by the position $y$ and the orientation $\varphi$. A narrow beam of an angular extent $\alpha $ makes a sector with an arclength ${\cal A}$; the ratio of the arclength to the angle is the radius of curvature, $\rho $. As the beam travels, both ${\cal A}$ and $\rho $ increase. Just before (after) a collision, we denote the radius of curvature by $\rho_{-}$ ($\rho_{+}$). By simple geometry, $\rho_{+}$ is given by
\begin{equation}\label{eq:polyref}
\rho_{+} = \rho_{-} \left( \frac{\sin \alpha }{\sin (\alpha - 2\theta )} \right) \left( \frac{1 - \nu \cos \theta - \nu \sin \theta \tan \xi }{1 - \nu \cos \theta + \nu \sin \theta \tan \xi } \right) 
\end{equation}
with $\nu = s_1/s_2$ and $\xi = (\alpha + \pi + 2\varphi - \theta )/2$. 

If $a$ is the ``radius" (half of the longest diagonal) of a scatterer, $s_1$ and $s_2$ can be derived in terms of $y, \varphi $ as
\begin{eqnarray}\label{eq:3}
s_1 &=& \frac{2 a\sin \varphi - y + \rho _- \sin (\alpha /2)}{\sin (\varphi + \theta /2)}, \nonumber \\
s_2 &=& \frac{-2 a\sin \varphi + y + \rho _-' \sin (\alpha /2)}{\sin (\varphi - \theta /2)}, \nonumber \\
\rho _-' &\approx & - \rho _- \cos (\theta - \alpha ).
\end{eqnarray}
Note that $y \sim a \sin \varphi $ \cite{note_a} (Fig. \ref{fig:fig3}). 
When the distribution of $s_1$ and $s_2$ is calculated, $y$ appears to be uniformly distributed on $[-a, a]$. 

If the beam hits multiple vertices, the collision can be re-cast in terms of striking across an equivalent vertex A, for different values of $\theta$, $s_1$ and $s_2$ as shown in Fig.~\ref{fig:fig3}.
In that case, $\theta$ can be written in terms of the number of vertices of the polygon $N$ and the number of vertices $n_v$ included in the rays as
\begin{equation}\label{eq:theta}
\theta = \pi - 2 \pi n_v / N .
\end{equation}
The bounding rays of the beam falling on the scatterer meet at what we call  the ``endpoints". The angle $\alpha $ is around the ratio of arclength between the endpoints and the angle of a section of the polygon. For small $\alpha, x$,  

\begin{equation}\label{eq:x}
x = N\frac{\alpha \rho_{-} }{a \cos \varphi }.
\end{equation}
\begin{equation}\label{eq:nv}
 n_v = \begin{cases}\texttt{Ceil}[x], & \text{with probability \texttt{frac}[x]}, \\   \texttt{Floor}[x], &  \text{otherwise}\end{cases}.
\end{equation}

We have assumed that it is only the ratio, $x$ that is important for deciding the number of vertices.
\begin{figure}[!]
\begin{center}
\includegraphics[scale=1.0,height=66mm]{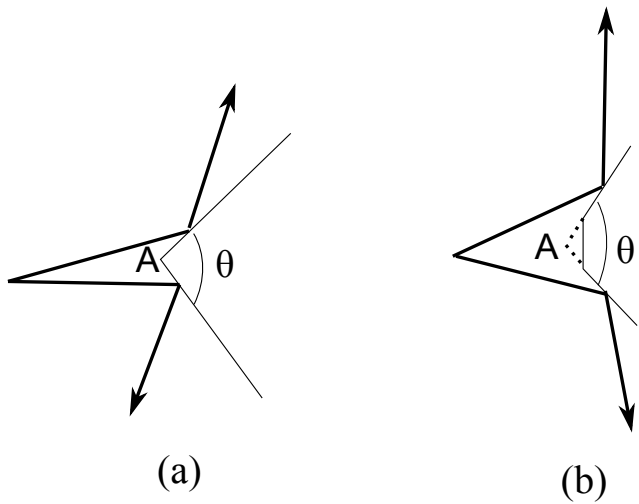}
\end{center}
\caption{(a) If the pencil of rays is incident (a) about a single vertex ($n_v = 1$), $\theta = \pi - 2\pi /N$. (b) If it is incident such that it includes one side of the scatterer ($n_v = 2$), then $\theta = \pi - 2\times2\pi /N$. }
\label{fig:fig4}
\end{figure}

\begin{figure}
\centering
\begin{minipage}{.5\textwidth}
  \centering
  \includegraphics[scale=0.6,height=33mm]{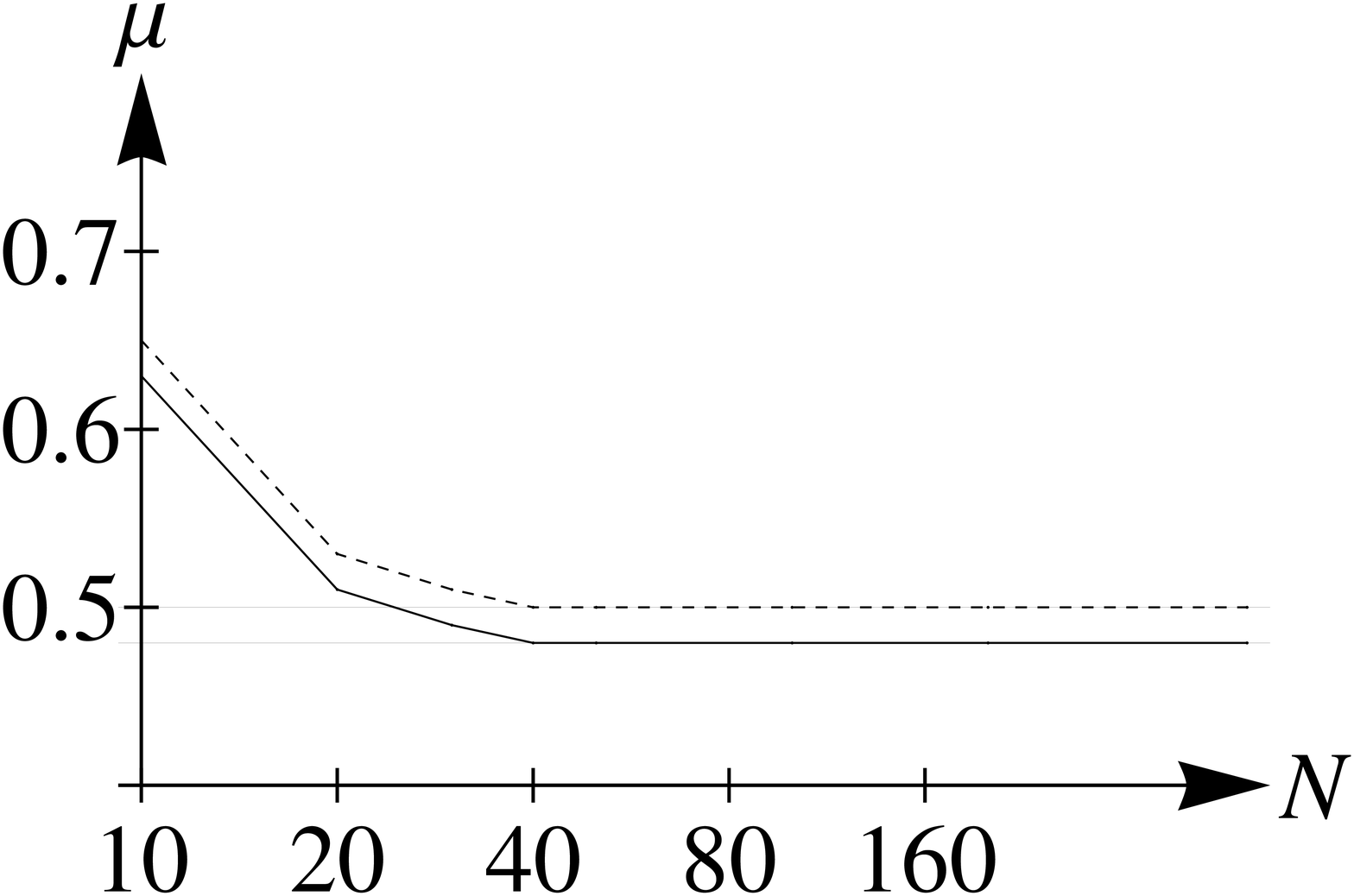}
\end{minipage}%
\begin{minipage}{.5\textwidth}
  \centering
  \includegraphics[scale=0.6,height=33mm]{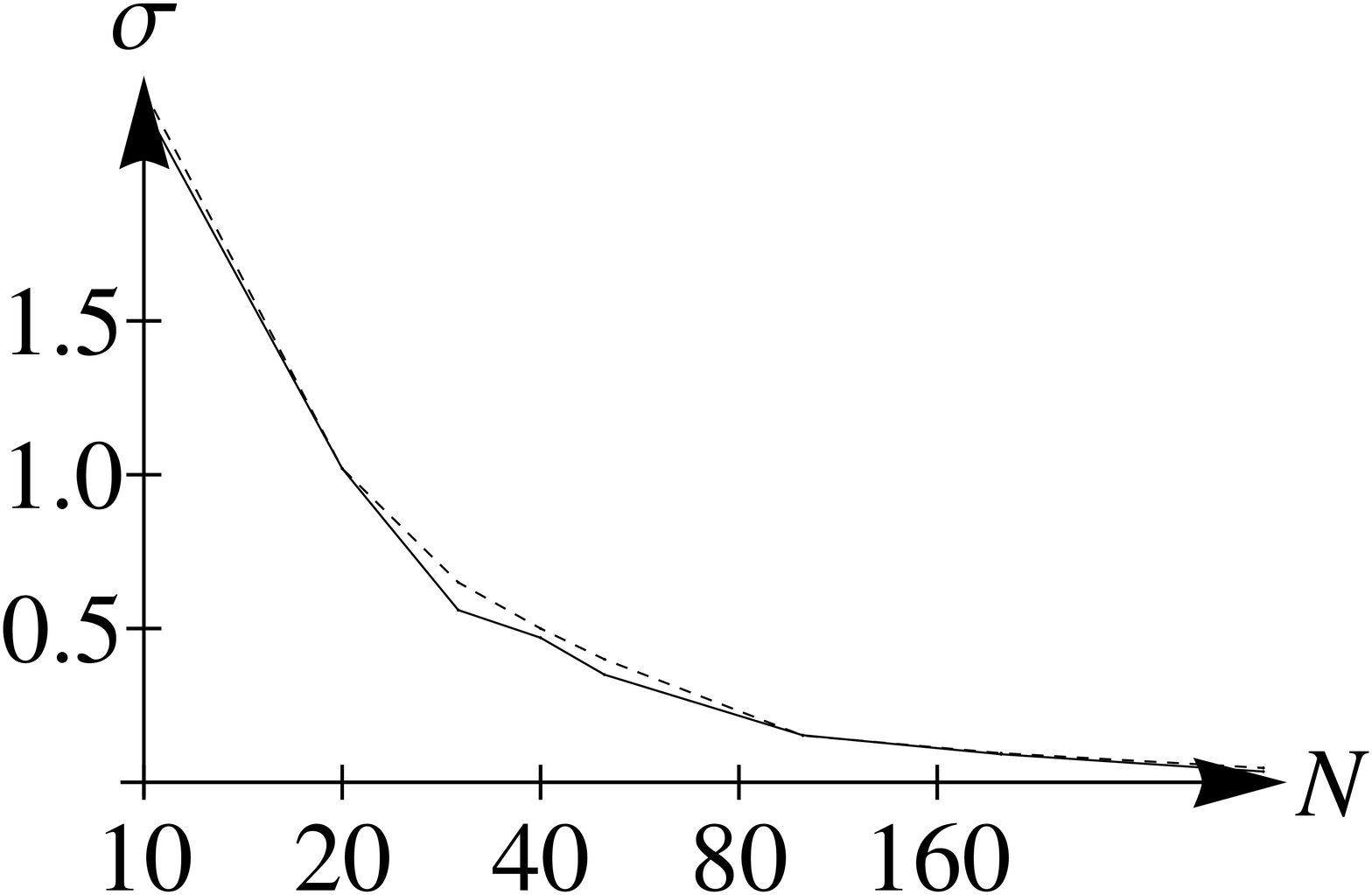}
\end{minipage}
\caption{The data for the distribution of $\rho_{+}$. The dashed line is for a large value of $\rho_{-} = 1000$ and the solid line is for a small value $\rho_{-} = 10$. The left panel shows the mode $\mu$. The horizontal lines indicate the expected answer using the spherical mirror reflection formula Eq.~(\ref{eq:circref}). The right panel shows the variance $\sigma$.}
\label{figmodvar}
\end{figure}

Substituting Eqs.~(\ref{eq:3}, \ref{eq:theta}) in Eq.~(\ref{eq:polyref}), and using a uniform distribution for $s_1$, $s_2$ and $\sin \varphi$, for a fixed $\rho_{-}$, we obtain a distribution for $\rho_+ /a$ that peaks at the value obtained using the mirror formula Eq.~(\ref{eq:circref}) for circular scatterers.
The mode and the variance of the the distribution obtained for $\rho_{+}$ have been plotted in Fig.~\ref{figmodvar}.
The graphs show that the variance $\sigma\rightarrow 0$ as $N\rightarrow\infty$ and the mode $\mu$ goes to a fixed value, the value obtained from (\ref{eq:circref}).
This establishes that Eq.~(\ref{eq:polyref}) is a polygonal generalization of the (circular) mirror formula Eq.~(\ref{eq:circref}).
This is an important result on which the subsequent discussion is based.

\section{Generalized Lyapunov exponent}

We briefly discuss the estimation of Lyapunov exponent for the Lorentz gas, following Sinai \cite{krylov-sinai,bob}.
This motivates the calculation of a finite-time exponent for the Ehrenfest gas.

\subsection{Circular scatterers}

In the case of random Lorentz gas, the separation of the trajectories after $i^{th}$ collision, $s_i$ can be iteratively written as 
\begin{equation}\label{eq:5}
s_i = s_{i-1} \left( \frac{\rho _{i-1} + v\tau _i}{\rho _{i-1}} \right)
\end{equation}
where $\tau _i$ is the time of flight between the $(i-1)^{th}$ and the $i^{th}$ collisions.
The Lyapunov exponent \cite{bob} is defined as the average rate of exponential separation of two close-by trajectories, whose initial separation $s_0$ and the initial angle between the trajectories $\alpha_0$ both go to $0$.
From the above equation, the Lyapunov exponent can be identified as \cite{bob}
\begin{eqnarray}\label{eq:7}
\lambda &=& \lim_{T \to \infty } \frac{1}{T} \sum_{i=1}^{n} \ln \left( 1 + \frac{v\tau _i}{\rho _{i-1}} \right) \nonumber \\
&=& \lim_{T \to \infty } \frac{v}{T} \int_{t_0}^{t_0+T} \frac{1}{\rho _{i-1} (t)} dt\nonumber\\
&=& v \bigg\langle\frac{1}{\rho _i\left(t\right)}\bigg\rangle
\end{eqnarray}
where $n$ is the number of collisions between $t_0$ and $t_0 + T$.
We can also write $\lambda$ as
\begin{equation}\label{eq:lya_basic}
\lambda = \frac{1}{\tau_f}\bigg\langle\ln\left(1+\frac{v \tau}{\rho_+}\right)\bigg\rangle .
\end{equation}
where $\tau_f$ is the mean free time between collisions and $\tau $ the time between subsequent collisions. Note that $(v\tau)$ (denoted by $\xi$) follows an exponential distribution with a mean equal to $\mu_f=v\tau_f$. In Eq.~(\ref{eq:lya_basic}), $\rho_{+}$ is a function of $\rho_{-}$ and $\varphi$ via the reflection formula Eq.~(\ref{eq:circref}). The average in Eq.~(\ref{eq:lya_basic}) is over the uniform distribution of $\sin \varphi$ and the exponential distribution of $\xi$:
\begin{equation}\label{eq:twoplus}
\lambda = \frac{1}{\tau_f}\bigg\langle\ln\left(2 + \frac{2 \xi}{a \cos\varphi}\right)\bigg\rangle_{\xi,\varphi}
\end{equation}
To obtain the average over $\xi $ and $\varphi$, we have
\begin{eqnarray}\label{eq:phiint}
\lambda &=& \frac{1}{\tau _f} \left[\frac{1}{2} \int_{-\pi /2}^{\pi /2} d\varphi \cos \varphi \left\{ \frac{1}{\mu _f} \int_{0}^{\infty} d\xi e^{-\xi /\mu _f} \log \left( 2 + \frac{2\xi }{a \cos \varphi } \right) \right\}  \right] \nonumber \\
&=& \frac{1}{2\tau _f} \left[\int_{-\pi /2}^{\pi /2} d\varphi \cos \varphi \left\{ -{\cal C} + \log \left( \frac{2\mu _f}{a \cos \varphi } \right) \right\} \right] \nonumber \\
&=& 2nav\left( 1 - {\cal C} - \log (2 n a^2) \right).
\end{eqnarray}
In the second step, $2$ is ignored in comparison with $2\xi /a\cos \varphi$ in the argument of the logarithm as it contributes significantly only at higher densities. The result we have is exactly the one  derived by van Beijeren and Dorfman \cite{henk_bob}, where ${\cal C}$ is the Euler-Mascheroni constant.




\subsection{Polygonal scatterers}

In order to bring out the subtleties in the Ehrenfest Gas, we first contrast it with the Lorentz Gas.
In the Lorentz gas, the angle of the beam $\alpha^{\prime}$ after reflection is given by
\begin{equation}\label{eq:angle_circle}
 \alpha^{\prime} = \alpha \left(\frac{\rho_-}{\rho_+}\right). 
\end{equation}
At each collision, the angle $\alpha$ is multiplied by a quantity $\rho_-/\rho_+$, which is independent of $\alpha$, as seen in Eq.~(\ref{eq:circref}).
However, for the Ehrenfest Gas, the angle of the beam after a collision with a polygonal scatterer is 
\begin{equation}\label{eq:angle_polygon}
 \alpha^{\prime} = 2 \pi - 2 \theta + \alpha.
\end{equation}
Once the beam strikes a vertex, $\theta \neq \pi$, $\alpha$ increases almost algebraically as $\theta $ is roughly proportional to $\alpha $. 
Since for the Lorentz Gas $\alpha$ increases multiplicatively, for any howsoever large time $T$, the initial angle can be chosen to make $\alpha '$ so small so as to ensure that the full beam hits the same {scatterers} with probability \emph{equal} to unity.
Hence the limit $T \rightarrow \infty$ can be taken in (\ref{eq:7}).
For the Ehrenfest gas, this is true only if we choose the initial angle of the beam so small that it will always strike a flat side without enclosing a vertex. This leads to a zero Lyapunov exponent, a well-known result.

\begin{figure}[!]
\begin{center}
\includegraphics[scale=1.0,height=77mm]{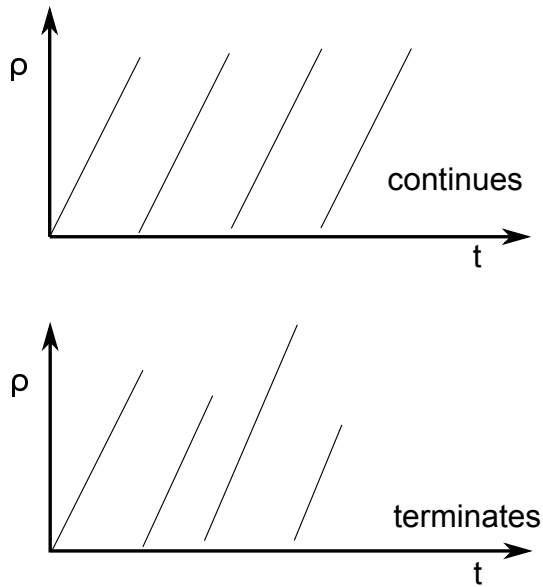}
\end{center}
\caption{The $\rho$ -- $t$ graph for the Lorentz gas and Ehrenfest Gas respectively. Note that this is only a schematic plot, assuming that all the distance between the scatterers is equal to the mean free path.}
\label{fig:stops}
\end{figure}

Beams with a finite size will eventually split in both the cases - Lorentz and Ehrenfest. However, as we want to capture the positive $\lambda $ of the Lorentz gas in the limit $N \to \infty$, we use a finite (and not infinitesimal) initial $\alpha $, which inevitably leads to splitting of the beam after a finite time. The splitting time $T$ could be longer if the size of the initial beam is made smaller. But then there will be more non-diverging collisions before hitting a vertex for the first time. Eqs (\ref{eq:circref},\ref{eq:polyref}) give the relations between the radii of curvature before and after a collision for Lorentz and Ehrenfest gases respectively. In a low density Lorentz gas, $\rho$ increases uniformly with time until it falls to about $(a \cos \varphi/2)$ after a collision, after which it increases uniformly again, as shown in Fig.~\ref{fig:stops}. The pattern continues as long as the full beam strikes the same scatterers. As observed above, since the beam can be made to strike the same scatterers up to an arbitrarily large time, the $\rho$--$t$ graph continues indefinitely.

However, for the Ehrenfest Gas where the beam may eventually enclose a vertex as the initial angle is not infinitesimal, the condition $\alpha '\mu_f \leq a$ is violated after a certain number of collisions, causing the $\rho$--$t$ graph to terminate, as in Fig.~\ref{fig:stops}. It is this time that we propose to employ for calculating the time-average in this very definition of the exponent. 

In analogy with the manner in which the Lyapunov exponent is calculated for  circular scatterers, as in Eq.~(\ref{eq:7}), we define the finite-time exponent for the polygonal scatterers by 
\begin{eqnarray}\label{eq:finitedef}
\lambda _f (T) &=& \frac{1}{\sum_{j=1}^{n} \tau _i}\sum_{i=1}^n \log  \left(1+ \frac{\xi _i}{\rho _i}\right) \nonumber \\
& \sim &\frac{1}{2T} \bigg\langle \sum_{i=1}^{n} \left( 1 - \log 2 + \log \frac{\xi _i}{a} \right) \bigg\rangle 
\end{eqnarray}
where with $\rho _i$ as $2a\cos \varphi _i$, the latter expression is obtained after averaging over the angle variable assuming that $\xi _i \gg \rho _i$.

The above expression is similar to the usual expression for the Lyapunov exponent, the difference being that we average over a finite time $T$. In general, $\lambda _f$ will depend on initial conditions, initial angle, and $T$. However, Ehrenfest dynamics is represented in terms of interval-exchange transformations (IET), where intervals correspond to the set of points on the disc from where scattering occurs. From collision-to-collision, there is a shuffling of the points belonging to the interval on the subsequent discs. This is the same as the dynamics inside a polygonal billiard \cite{veech,hannay,jain}. And recently, it has been proved that the dynamics induced by IET is weakly mixing \cite{avila,ferenczi}. Since weak mixing implies ergodicity \cite{cfs22}, initial conditions will not matter over long times.   

For further evaluation, we need a distribution function that takes into account the different lengths of the ``teeth"  in the $\rho$--$t$ graph (Fig.~\ref{fig:stops}) for the Ehrenfest Gas. Note that a new tooth begins only when $\alpha$ and $\rho$ change upon a collision. For a circular scatterer, they change at each collision and thus the length of each tooth follows the distribution of times taken along the free paths between the scatterers.
However, for the polygonal scatterer, this is not true of every collision.
For the first few collisions, it is possible that the rays do not hit across a vertex, and hence face $\theta=\pi$, keeping $\alpha$ constant according to Eq.~(\ref{eq:angle_polygon}).
This aspect changes the value of $\rho_i$ in Eq.~(\ref{eq:finitedef}), and thus the value of the finite-time Lyapunov exponent if the initial alpha is chosen too small (or, very rarely, if one of the first few free times is much shorter than the average). For large $N$ and $\alpha_0 v t_0 > a/N$, usually all collisions will be diverging. The change in finite time Lyapunov exponent will therefore occur for very small $\alpha_0$ only. 

It is hence crucial to determine the number of collisions on straight edge before encountering a vertex. A simulation was performed on \textsc{Mathematica} using the reflection formula (\ref{eq:polyref}), for an Ehrenfest Gas with a mean free path of $8$ in units of the size of the scatterers. Scatterers with $N = 10^5$ sides were chosen, with an initial angle between nearby rays taken as $ \alpha_0 = 3 \times 10^{-5}$ radians. This resulted in the applicable time scale to be as large as around 100 units (of $v/a$ where $v$ is the velocity of the wind and $a$ is the measure of the size of a scatterer). The number of teeth ranged up to 4. 

Denoting the probability $p$ of hitting the vertex in the first collision, we note that its approximate value can be calculated from the 
following equation:
\begin{eqnarray}\label{eq:prob}
p&=& I\left(\frac{\mu_f
\alpha}{2a\sin(\pi/N)}\right); \nonumber \\
\mbox{where~} I(x) &=& x \mbox{~for~} x < 1, \mbox{~and~} 1 \mbox{otherwise}. 
\end{eqnarray}
It is a measure of $N$, and as $N \rightarrow \infty$, $p \rightarrow 1$. 

\begin{figure}
\centering
\begin{minipage}{.5\textwidth}
  \centering
  \includegraphics[scale=0.6,height=33mm]{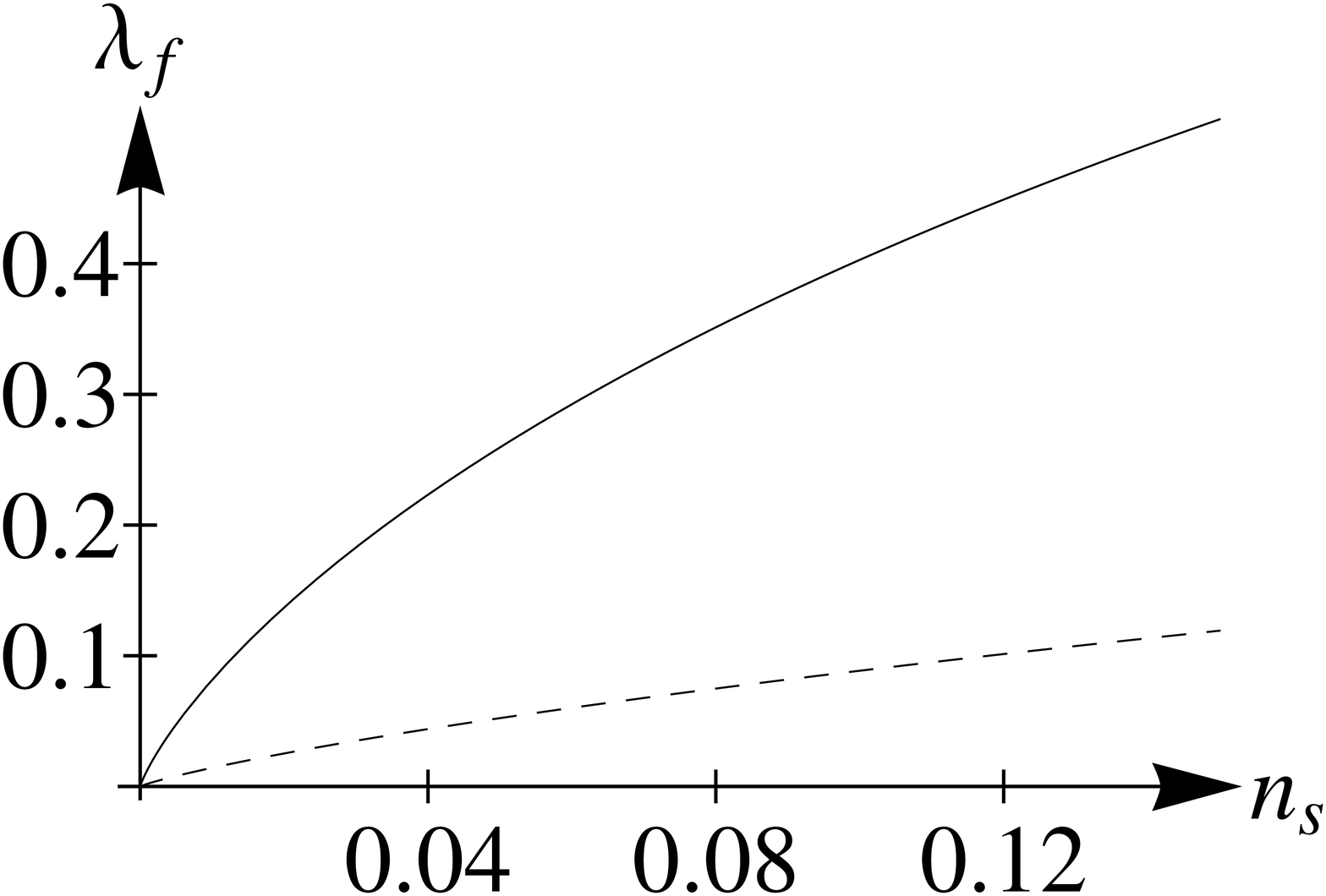}
\end{minipage}%
\begin{minipage}{.5\textwidth}
  \centering
  \includegraphics[scale=0.6,height=33mm]{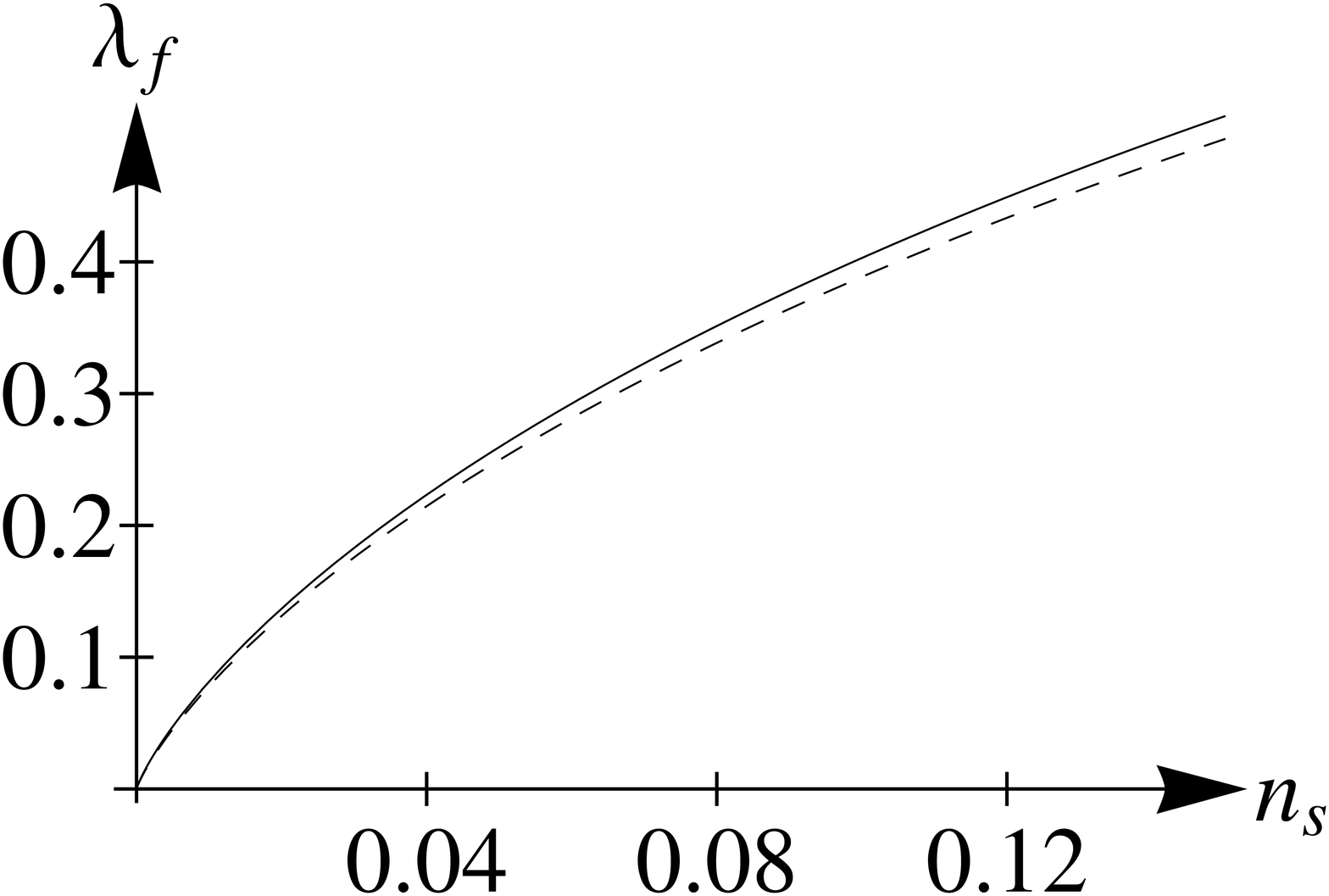}
\end{minipage}
\caption{The solid curve corresponds to the usual Lyapunov exponent for the Lorentz gas and the dashed curve corresponds to that for the finite-time exponent of the Ehrenfest gas, both plotted as a function of the scatterer density. The left panel corresponds to an initial $p=0.01$ and the right panel corresponds to $p=0.9$.}
\label{fig:ftlyapunov}
\end{figure}

We present the  results for $\lambda_f$ as calculated for the first tooth in Fig.~\ref{fig:ftlyapunov}. If we set the mean free path to 16 times the radius of the scatterer, and the initial angle $\alpha=3\times10^{-4}$, then the case of $p=0.01$ corresponds to a scatterer with 12 sides, and $p=0.9$ corresponds to a scatterer with around 1200 sides. As the number of sides of the scatterer increases, and hence $p \to 1$, the finite-time Lyapunov exponent $\lambda_f$ for the Ehrenfest gas approaches the usual Lyapunov exponent $\lambda$ for the Lorentz gas.
The time of validity of the exponent can
be estimated as follows: the condition to be met is $\alpha _0 e^{\lambda t} \approx 2a$. With $\alpha _0 \approx 1/N$, we see that $t$ is proportional to $\log N$. 

\begin{figure}[!]
\begin{center}
\includegraphics[scale=1.0,height=66mm]{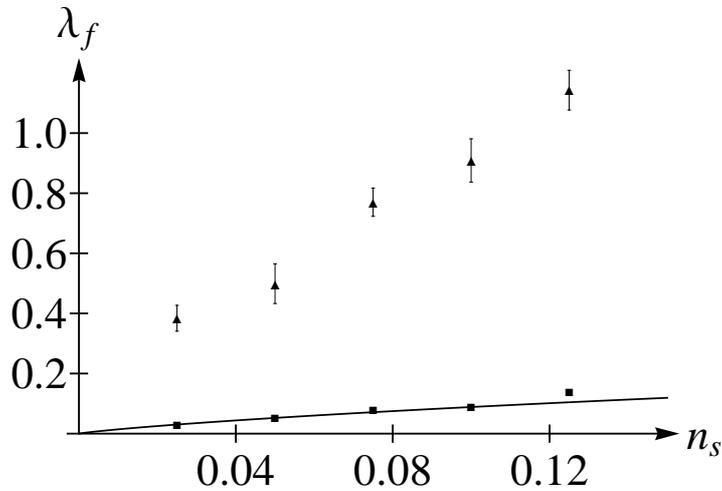}
\end{center}
\caption{The solid curve is the analytic expression for the Lyapunov exponent calculated over single tooth. The square points represent the simulation results for the same. Note that the error bars for these points are negligible at this scale. The triangular points represent the Lyapunov exponent as calculated with averages over multiple teeth. This graph contains data for $p=0.01$.}
\label{fig:simul}
\end{figure}

The exponent calculated over a single tooth gives a lower bound on the more accurate exponent calculated by an average over multiple teeth, which will be over a total duration of the order $\log N$. The point of this is to show that a strictly non-zero finite-time exponent does exist. We have verified this in a simulation on \textsc{Mathematica}, presented in Fig.~\ref{fig:simul}.

\section{Summary}

We have studied the case in the theory of dynamical systems where beam splitting leads to non-integrability. Although the Lyapunov exponent of such systems is known to be zero, we may define a finite-time exponent which is positive. This is particularly meaningful for polygonal scatterers with large number of sides.

The reflection law for the case of a regular polygonal scatterer was found.  More precisely, Eq. (\ref{eq:polyref}) holds for any convex scatterer provided the two intersecting segments are replaced by the tangents to the scatterer.  Having found this, we have argued that for pseudointegrable systems (which are non-chaotic), it is sometimes possible to define a finite-time exponent. In the limit of the number of vertices becoming infinite (and hence the scatterer becoming circular), the exponent found here approaches the Lyapunov exponent  for the Lorentz gas. Thus, on both counts - the collision rule and Lyapunov exponent - we have generalized the results. We have also argued that the finite-time exponent is valid for times of the order $\log N$.

\section*{Acknowledgements}   
SM and SC would like to thank the National Initiative for Undergraduate Sciences program under the Homi Bhabha Centre for Science Education (HBCSE) for the opportunity and funding, and, Bhabha Atomic Research Centre (BARC) for the facilities for carrying out this work. The authors also thank the Referee for very valuable suggestions.


\begin{thebibliography}{99}
\bibitem{footnote} Integrability in the senes of Liouville-Arnol'd refers to existence of $f$ functionally independent constants of the motion that are in involution and where the vector fields in phase space are smooth everywhere for a system with $f$ degrees of freedom. Non-integrability refers to all the behaviours which result from any of the abovementioned conditions to be violated. Thus, defocussing scatterers will result in chaotic dynamics but scatterers with straight edges will give a non-integrable dynamics due to vector fields being singular at vertices \cite{arnold78,jain}.
\bibitem{note_a} For a polygonal scatterer, $y$ will be a uniform distributed around $a \sin \varphi $ with the limits depending on $\varphi $. The values it can take also depend on $N$. As $N \to \infty $, the possible values of $y$ will be strongly peaked around $\sin \varphi $.
\end{thebibliography}
\end{document}